\documentstyle[12pt,epsfig]{article}

\newcommand{\alt}{\,\rlap{\lower4.5pt\hbox{$\mathchar\sim$}}\raise2.5pt
\hbox{$<$}\,}
\newcommand{\re}{\mathop{\mbox{Re}}\nolimits}
\newcommand{\im}{\mathop{\mbox{Im}}\nolimits}
\newcommand{\cl}{\mathop{{\mbox{Cl}}_2}\nolimits}

\catcode`@=11
\newcount\@tempcntc
\def\@citex[#1]#2{\if@filesw\immediate\write\@auxout{\string\citation{#2}}\fi
  \@tempcnta\z@\@tempcntb\m@ne\def\@citea{}\@cite{\@for\@citeb:=#2\do
    {\@ifundefined
       {b@\@citeb}{\@citeo\@tempcntb\m@ne\@citea\def\@citea{,}{\bf
?}\@warning
       {Citation `\@citeb' on page \thepage \space undefined}}%
    {\setbox\z@\hbox{\global\@tempcntc0\csname b@\@citeb\endcsname\relax}%
     \ifnum\@tempcntc=\z@ \@citeo\@tempcntb\m@ne
       \@citea\def\@citea{,}\hbox{\csname b@\@citeb\endcsname}%
     \else
      \advance\@tempcntb\@ne
      \ifnum\@tempcntb=\@tempcntc
      \else\advance\@tempcntb\m@ne\@citeo
      \@tempcnta\@tempcntc\@tempcntb\@tempcntc\fi\fi}}\@citeo}{#1}}
\def\@citeo{\ifnum\@tempcnta>\@tempcntb\else\@citea\def\@citea{,}%
  \ifnum\@tempcnta=\@tempcntb\the\@tempcnta\else
   {\advance\@tempcnta\@ne\ifnum\@tempcnta=\@tempcntb \else
\def\@citea{--}\fi
    \advance\@tempcnta\m@ne\the\@tempcnta\@citea\the\@tempcntb}\fi\fi}
\catcode`@=12

\begin{document}
\title{\vskip-3cm{\baselineskip14pt
\centerline{\normalsize DESY 01-203\hfill ISSN 0418-9833}
\centerline{\normalsize hep-ph/0112023\hfill}
\centerline{\normalsize November 2001\hfill}}
\vskip1.5cm
Theoretical Aspects of Standard-Model Higgs-Boson Physics at a Future $e^+e^-$
Linear Collider}
\author{{\sc Bernd A. Kniehl}\\
{\normalsize II. Institut f\"ur Theoretische Physik, Universit\"at Hamburg,}\\
{\normalsize Luruper Chaussee 149, 22761 Hamburg, Germany}}

\date{}

\maketitle

\thispagestyle{empty}

\begin{abstract}
The Higgs boson is the missing link of the Standard Model of elementary
particle physics.
We review its decay properties and production mechanisms at a future $e^+e^-$
linear collider and its $e^-e^-$, $e^\pm\gamma$, and $\gamma\gamma$ modes,
with special emphasis on the influence of quantum corrections.
We also discuss how its quantum numbers and couplings can be extracted from 
the study of appropriate final states.
\end{abstract}

\newpage

\section{\label{sec:one}Introduction}

The SU(2)$_I\times$U(1)$_Y$ structure of the electroweak interactions has been
consolidated by an enormous wealth of experimental data during the past three
decades.
The canonical way to generate masses for the fermions and intermediate bosons
without violating this gauge symmetry in the Lagrangian is by the Higgs
mechanism of spontaneous symmetry breaking.
In the minimal standard model (SM), this is achieved by introducing one
complex SU(2)$_I$-doublet scalar field $\Phi$ with $Y=1$.
The three massless Goldstone bosons which emerge via the electroweak symmetry
breaking are eaten up to become the longitudinal degrees of freedom of the
$W^\pm$ and $Z$ bosons, {\em i.e.}, to generate their masses, while one
$CP$-even Higgs scalar boson $H$ remains in the physical spectrum.
The Higgs potential $V$ contains one mass and one self-coupling.
Since the vacuum expectation value is fixed by the relation
$v=2^{-1/4}G_F^{-1/2}\approx246$~GeV, where $G_F$ is Fermi's constant, there
remains one free parameter in the Higgs sector, namely $M_H$.
In fact, one has
\begin{equation}
V=\lambda H^2\left(v+\frac{H}{2}\right)^2+\cdots,
\end{equation}
where $\lambda=M_H^2/(2v^2)$.
The Higgs boson has the quantum numbers of the vacuum, namely electric
charge $Q=0$, spin, parity, and charge conjugation $J^{PC}=0^{++}$.
It has tree-level couplings to all massive particles with strengths that are
determined by their masses, viz.\ $g_{ffH}=M_f/v$, $g_{VVH}=2M_V^2/v$,
$g_{VVHH}=2M_V^2/v^2$, $g_{HHH}=6v\lambda$, and $g_{HHHH}=6\lambda$, where $f$
denotes a generic fermion and $V=W,Z$.
At a future $e^+e^-$ linear collider (LC), an important experimental task will
be to determine of the Higgs quantum numbers and couplings in order to
distinguish between the minimal SM and possible extensions.
In particular, the measurement of the Higgs self-couplings will allow one to
directly test the Higgs mechanism.

\begin{figure}[ht]
\begin{center}
\begin{tabular}{cc}
\parbox{7.5cm}{
\epsfig{file=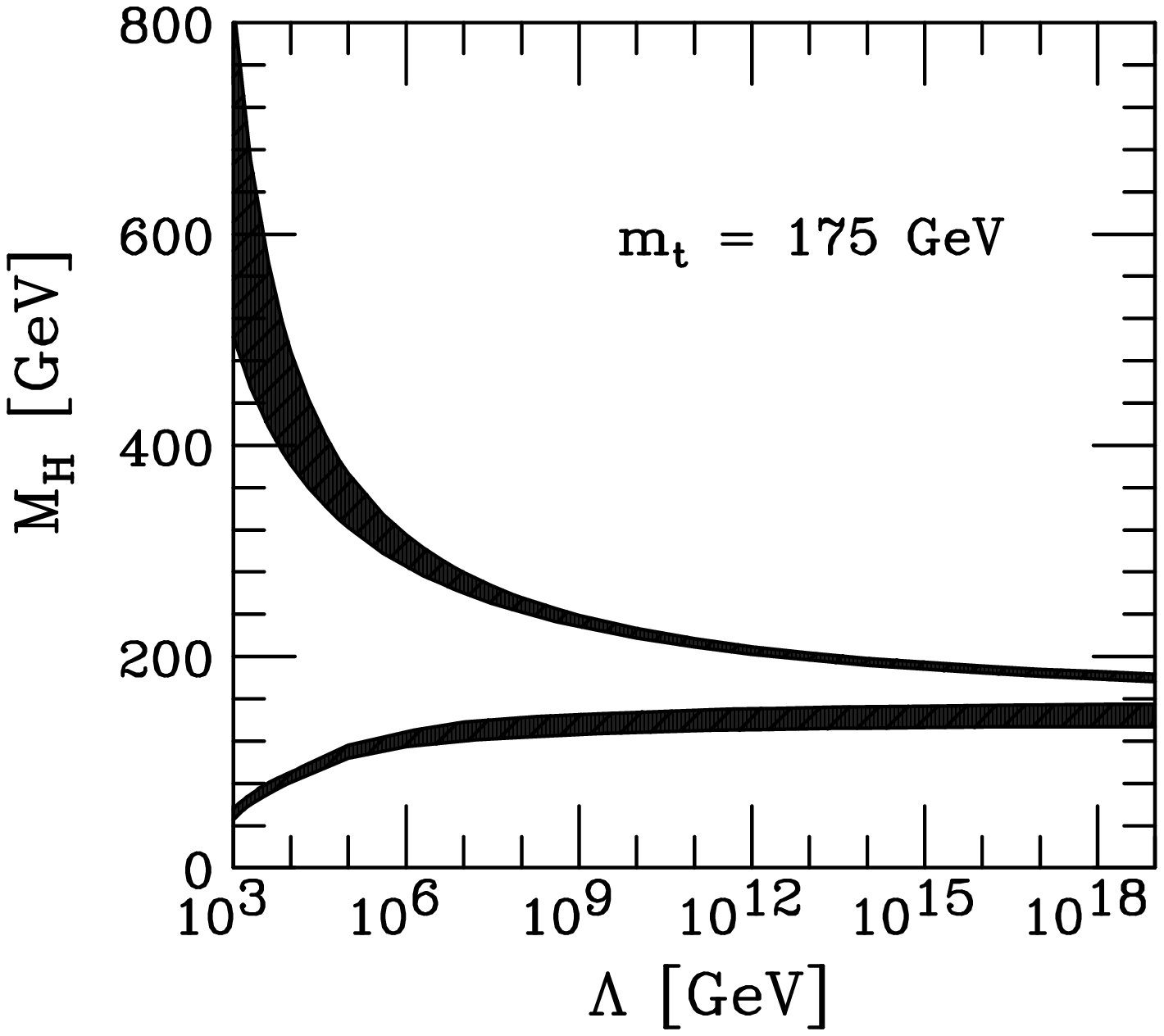,width=7.5cm,bbllx=94pt,bblly=356pt,bburx=502pt,%
bbury=720pt}}
&
\parbox{7.5cm}{
\epsfig{file=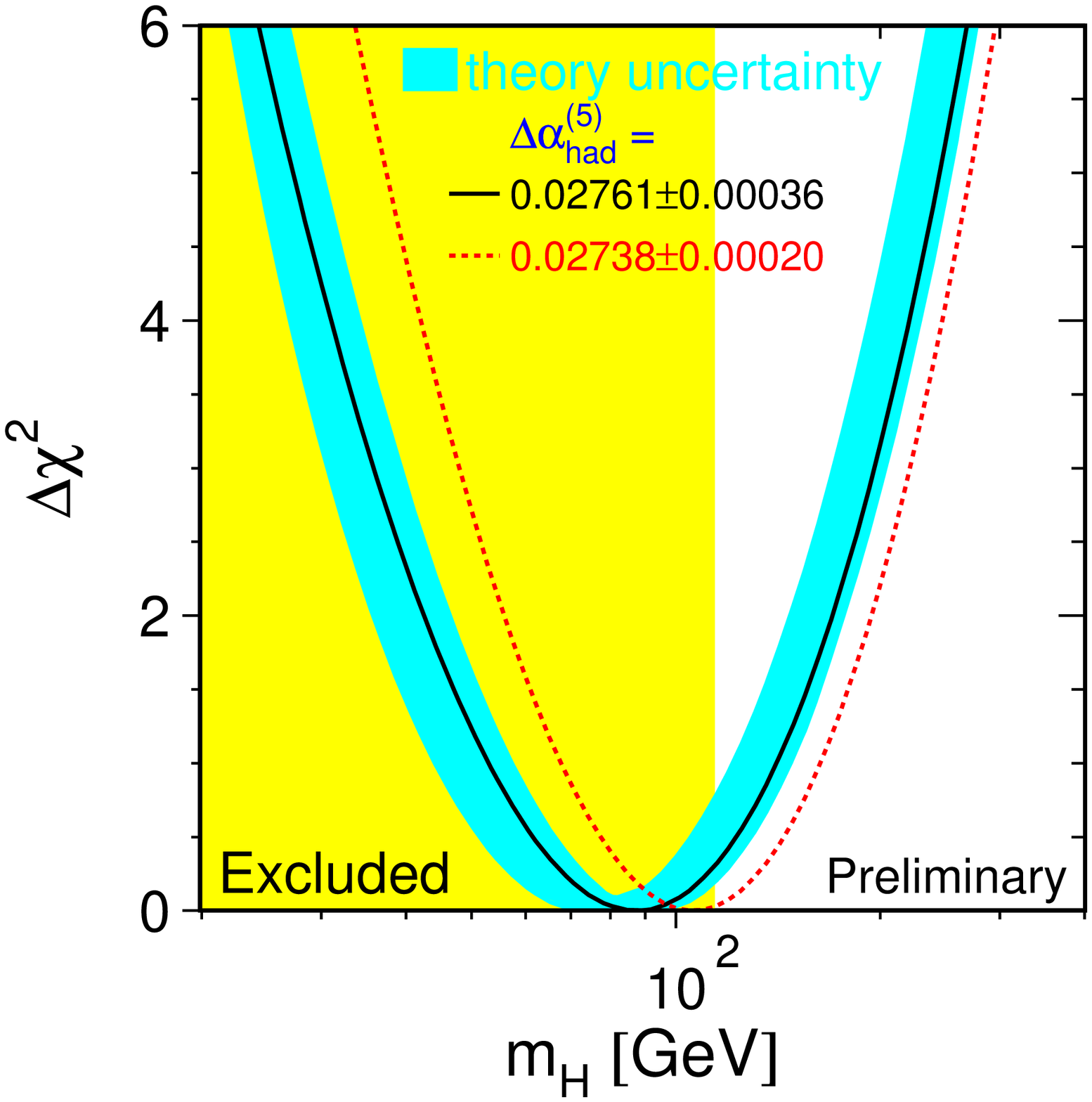,width=7.5cm,bbllx=17pt,bblly=36pt,bburx=542pt,%
bbury=565pt}}\\
(a) & (b)
\end{tabular}
\caption{(a) Triviality and vacuum-stability bounds on $M_H$
\protect\cite{ham} and (b) $\Delta\chi^2=\chi^2-\chi_{\rm min}^2$ as a
function of $M_H$ from fits to electroweak precision data from LEP taking into
account the direct determinations of $M_W$ and $M_t$ \protect\cite{ewwg}.}
\label{fig:blue}
\end{center}
\end{figure}

Roughly speaking, the requirement that the running Higgs self-coupling
$\lambda(\mu)$, where $\mu$ is the renormalization scale, stays finite
(positive) for all values $\mu<\Lambda$, where $\Lambda$ is the cutoff beyond
which new physics operates, leads to the triviality upper bound
(vacuum-stability lower bound) on $M_H$ \cite{cab}.
Assuming the SM to be valid up to the grand-unified-theory scale
$\Lambda\approx10^{16}$~GeV, one thus obtains $130\alt M_H\alt185$~GeV
\cite{ham} [see Fig.~\ref{fig:blue}(a)].
This range comfortably lies between the lower bound on $M_H$ from direct
searches at CERN LEP2, 113~GeV, and the 95\% confidence level upper bound from
electroweak precision tests \cite{ewwg}, 212~GeV, based on
$\Delta\alpha_{\rm had}^{(5)}(M_Z^2)=0.02738\pm0.00020$ \cite{mar}, and it is
compatible with the $1\sigma$ range $76<M_H<181$~GeV resulting from the latter
\cite{ewwg} [see Fig.~\ref{fig:blue}(b)].

It is interesting to consider a hypothetical scenario in which the Higgs boson
is absent and to constrain the mass scale $\Lambda$ of the new physics that
would take its place.
Using recent measurements of $\sin^2\theta_{\mathrm{eff}}^{\mathrm{lept}}$ and
$M_W$ \cite{ewwg}, one finds that, in a class of theories characterized by
simple conditions, the upper bound on $\Lambda$ is close to or smaller than
the upper bound on $M_H$, while in the complementary class $\Lambda$ is not
restricted by such considerations \cite{kni}.

This review is organized as follows.
In Sects.~\ref{sec:two} and \ref{sec:three}, we discuss the decay properties
of the Higgs boson and its main production mechanisms in $e^+e^-$, $e^-e^-$,
$e^\pm\gamma$, and $\gamma\gamma$ collisions, emphasizing the influence of
radiative corrections.
In Sect.~\ref{sec:four}, we explain how to extract its quantum numbers and
couplings from the study of final states.
Sect.~\ref{sec:five} contains our conclusions and a brief outlook.

\section{\label{sec:two}Decay properties}

At the tree level, the Higgs boson decays to pairs of massive fermions and 
gauge boson, the partial widths being
\begin{eqnarray} 
\Gamma\left(H\to f\bar f\right)&=&\frac{g_{ffH}^2}{4\pi}\,\frac{N_cM_H}{2} 
\left(1-\frac{1}{r_f}\right)^{3/2},
\nonumber\\
\Gamma(H\to VV)&=&\frac{g_{VVH}^2}{4\pi}\,\frac{3\delta_V}{8M_H}
\left(1-\frac{4}{3}r_V+\frac{4}{3}r_V^2\right)
\left(1-\frac{1}{r_V}\right)^{1/2},
\end{eqnarray}
respectively, where $N_c=1$ (3) for leptons (quarks), $\delta_{W,Z}=2,1$, and
$r_i=M_H^2/(4M_i^2)$.
If $1/4<r_i<1$ ($r_i<1/4$), then one of the (both) final-state particles are
forced to be off shell, so that one is dealing with three-particle
(four-particle) decays \cite{hzgg}.
The Higgs boson also couples to photons (gluons), through loops involving
charged (coloured) massive particles, and one is led to consider the
loop-induced decays $H\to Z\gamma$ \cite{HZp}, $H\to\gamma\gamma$ \cite{Hpp},
$H\to gg$ \cite{Hgg}, {\it etc.}

In order to match the high experimental precision to be achieved with a future
$e^+e^-$ LC, it is indispensable to take radiative corrections into account.
A review of radiative corrections relevant for SM Higgs-boson phenomenology
may be found in Refs.~\cite{pr,fp}. 
At one loop, the electroweak corrections to $\Gamma\left(H\to f\bar f\right)$
\cite{fle,hff}, $\Gamma(H\to VV)$ \cite{fle,hzz,hww}, and
$\Gamma\left(H\to Zf\bar f\right)$ \cite{pr} and the QCD ones to
$\Gamma(H\to q\bar q)$ \cite{hqq} are well established, including the
dependence on all particle masses.
Beyond one loop, only dominant classes of corrections were investigated,
sometimes only in limiting cases.
These include corrections enhanced by the strong-coupling constant $\alpha_s$,
the top Yukawa coupling $g_{ttH}$, and the Higgs self-coupling $\lambda$.
Specifically, the two-loop QCD corrections were found for
$\Gamma(H\to l^+l^-)$ \cite{hll}, $\Gamma(H\to q\bar q)$ ($q\ne t$)
\cite{gor,lar}, $\Gamma\left(H\to t\bar t\,\right)$ \cite{har},
$\Gamma(H\to Z\gamma)$ \cite{spi}, $\Gamma(H\to\gamma\gamma)$ \cite{zhe},
and $\Gamma(H\to gg)$ \cite{lar,ina}.
Even three-loop QCD corrections were calculated, namely for
$\Gamma(H\to q\bar q)$ ($q\ne t$) \cite{che}, $\Gamma(H\to\gamma\gamma)$
\cite{ste}, and $\Gamma(H\to gg)$ \cite{cks}.
In the last case, they are quite significant, the correction factor being
\cite{cks}
\begin{equation}
K_{gg}=1+\frac{215}{12}\,\frac{\alpha_s^{(5)}(M_H)}{\pi}
+\left(\frac{\alpha_s^{(5)}(M_H)}{\pi}\right)^2
\left(156.808-5.708\,\ln\frac{M_t^2}{M_H^2}\right),
\end{equation}
which approximately amounts to $1+0.66+0.21$ for $M_H=100$~GeV.

An efficient way of obtaining corrections leading in
$X_t=g_{ttH}^2/(4\pi)^2$ to processes involving low-mass Higgs bosons is to
construct an effective Lagrangian by integrating out the top quark.
This may be conveniently achieved by means of a low-energy theorem \cite{let},
which relates the amplitudes of two processes which differ by the insertion of
an external Higgs-boson line carrying zero four-momentum.
A na\"\i ve version of it may be derived by observing the following two
points:
(i) the interactions of the Higgs boson with the massive particles in the SM
emerge from their mass terms by substituting $M_i\to M_i(1+H/v)$; and
(ii) a Higgs boson with zero four-momentum is represented by a constant 
field.
This immediately implies that a zero-momentum Higgs boson may be attached
to an amplitude, ${\cal M}(A\to B)$, by carrying out the operation
\begin{equation}
\label{let}
\lim_{p_H\to0}{\cal M}(A\to B+H)={1\over v}\sum_i
{M_i\partial\over\partial M_i}{\cal M}(A\to B),
\end{equation}
where $i$ runs over all massive particles which are involved in the transition
$A\to B$.
This low-energy theorem comes with two caveats:
(i) the differential operator in Eq.~(\ref{let}) does not act on the $M_i$
appearing in coupling constants, since this would generate tree-level vertices
involving the Higgs boson that do not exist in the SM; and
(ii) Eq.~(\ref{let}) must be formulated for bare quantities if it is to be
applied beyond the leading order.

In this way, the effective Lagrangian describing the $l^+l^-H$, $W^+W^-H$, and
$ZZH$ interactions is found to be
\begin{equation}
{\cal L}_{\rm eff}=\frac{H}{v}\left[-\sum_lm_l\bar ll(1+\delta_u)
+2M_W^2W_\mu^-W^{+\mu}(1+\delta_{WWH})
+M_Z^2Z_\mu Z^\mu(1+\delta_{ZZH})\right],
\label{eq:eff}
\end{equation}
with \cite{hll,was,fer,gam}
\begin{eqnarray}
\delta_u&=&X_t\left\{\frac{7}{2}+3\left[\frac{149}{8}-6\zeta(2)\right]X_t
-[3+2\zeta(2)]A-56.703\,A^2\right\},
\nonumber\\
\delta_{WWH}&=&X_t\left\{-\frac{5}{2}+\left[\frac{39}{8}-18\zeta(2)\right]X_t
+[9-2\zeta(2)]A+27.041\,A^2\right\},
\nonumber\\
\delta_{ZZH}&=&X_t\left\{-\frac{5}{2}-\left[\frac{177}{8}+18\zeta(2)\right]X_t
+[15-2\zeta(2)]A+17.117\,A^2\right\},
\label{eq:del}
\end{eqnarray}
where $\zeta$ is Riemann's zeta function, with value $\zeta(2)=\pi^2/6$, and
$A=\alpha_s^{(6)}(M_t)/\pi$.
Notice that $\delta_u$ is universal in the sense that it comprises just the
renormalizations of the Higgs-boson wave function and vacuum expectation 
value.
The analytic expressions of the $O(A^2)$ terms may be found in 
Ref.~\cite{fer}.
In $O\left(X_t^2\right)$, also the full $M_b$ dependence is available
\cite{gam}.
From Eq.~(\ref{eq:eff}), one reads off that $\Gamma(H\to l^+l^-)$,
$\Gamma(H\to W^+W^-)$, and $\Gamma(H\to ZZ)$ receive the correction factors
\begin{eqnarray}
K_{ll}&=&(1+\delta_u)^2,
\nonumber\\
K_{WW}&=&(1+\delta_{WWH})^2,
\nonumber\\
K_{ZZ}&=&(1+\delta_{ZZH})^2,
\end{eqnarray} 
respectively.
The $O(X_t)$, $O\left(X_t^2\right)$, and $O(X_tA)$ corrections to
$\Gamma(H\to q\bar q)$, where $q\ne b,t$, coincide with those for
$\Gamma(H\to l^+l^-)$.
The $O(X_tA^2)$ corrections to $\Gamma(H\to q\bar q)$ were found in
Ref.~\cite{cks1}.
The effective-Lagrangian method in connection with the low-energy theorem was
also employed to obtain the $O(X_tA)$ \cite{hbb} and $O(X_tA^2)$ \cite{cks1}
corrections to $\Gamma(H\to b\bar b)$, the $O(X_t)$ corrections to
$\Gamma(H\to\gamma\gamma)$ \cite{gam}, and the $O(X_t)$ \cite{gam,cks1,dg} and
$O(X_tA)$ \cite{mat} corrections to $\Gamma(H\to gg)$.

The expansion of $\Delta\rho=1-1/\rho$, which measures the deviation of the
electroweak $\rho$ parameter from unity, analogous to Eq.~(\ref{eq:del}) reads
\cite{djo,avd,bij}
\begin{equation}
\Delta\rho=X_t\{3+3[19-12\zeta(2)]X_t-2[1+2\zeta(2)]A-43.782\,A^2\}.
\label{eq:rho}
\end{equation}
The analytic expression of the $O(A^2)$ term and the full $M_b$ dependence in
$O\left(X_t^2\right)$ may be found in Refs.~\cite{avd,bij}, respectively.
The $O\left(X_t^2\right)$ term exhibits a strong dependence on $M_H$, so that
its value for $M_H=0$ does not provide a useful approximation for realistic
values of $M_H$ \cite{bar}.
Furthermore, subleading electroweak two-loop corrections, of
$O\left(X_tG_FM_W^2\right)$, for $e\nu_\mu$ scattering and muon decay are not
actually suppressed in magnitude against the $O\left(X_t^2\right)$ one
\cite{deg}.
Thus, the approximations by the $O\left(X_t^2\right)$ terms for $M_H=0$ in
Eq.~(\ref{eq:del}) should be taken with a grain of salt.
The coefficients of $X_tA$ and $X_tA^2$ in Eqs.~(\ref{eq:del}) and
(\ref{eq:rho}) are all negative and sizeable relative to the one of $X_t$.
This is related to the use of the pole mass $M_t$.
In fact, the convergence behaviour of these expansions may be considerably
improved \cite{fer} by expressing them in terms of the scale-invariant
$\overline{\rm MS}$ mass, $\mu_t=m_t(\mu_t)$, which is related to $M_t$ by
\cite{gra}
\begin{equation}
\mu_t=M_t\left(1-\frac{4}{3}A-6.459\,A^2\right).
\end{equation}

\begin{figure}[ht]
\begin{center}
\centerline{\epsfig{figure=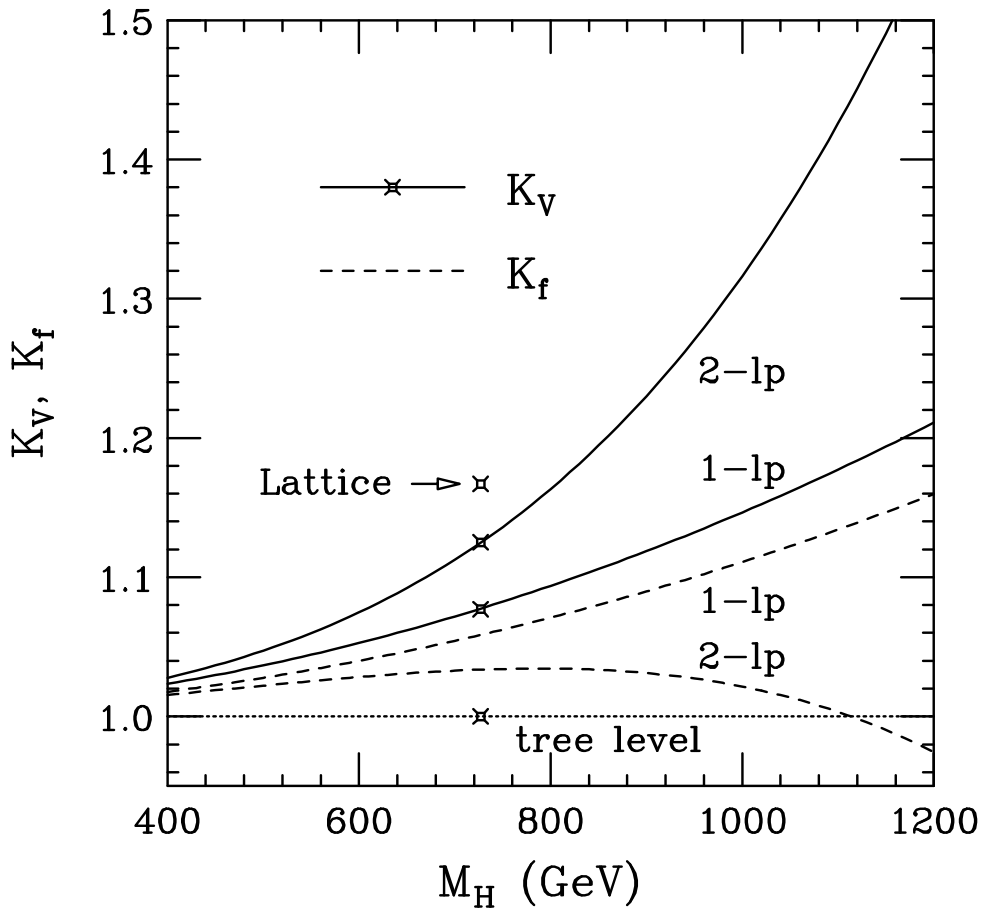,width=12cm,angle=90}}
\caption{$K_{VV}$ and $K_{ff}$ to $O(\lambda)$ and $O(\lambda^2)$ as functions
of $M_H$ \protect\cite{fri}.
The crosses indicate the tree-level, one-loop, two-loop, and nonperturbative
values of $K_{VV}$ at $M_H=727$~GeV.}
\label{fig:frink}
\end{center}
\end{figure}

The electroweak corrections for processes involving high-mass Higgs bosons,
with $M_H\gg2M_Z$, are dominated by powers of the Higgs self-coupling
$\lambda$.
These terms may be conveniently obtained by applying the Goldstone-boson
equivalence theorem \cite{cor}.
This theorem states that the leading high-$M_H$ electroweak contribution to a
Feynman diagram may be calculated by replacing the intermediate bosons $W^\pm$
and $Z$ with the respective would-be Goldstone bosons $w^\pm$ and $z$ of the
symmetry-breaking sector.
In this limit, the gauge and Yukawa couplings may be neglected against 
$\lambda$.
By the same token, the Goldstone bosons may be taken to be massless, and the
fermion loops may be omitted.
In this way, $\Gamma\left(H\to f\bar f\,\right)$ \cite{vel,dur},
$\Gamma(H\to W^+W^-)$, and $\Gamma(H\to ZZ)$ \cite{wil,ghi,fri} were studied
through $O(\lambda^2)$.
The resulting correction factor $K_{ff}$ for
$\Gamma\left(H\to f\bar f\,\right)$ is independent of the fermion flavour $f$.
Similarly, $\Gamma(H\to W^+W^-)$ and $\Gamma(H\to ZZ)$ receive the same 
correction factor $K_{VV}$.
In the on-mass-shell (OS) renormalization scheme, the results read
\cite{vel,dur,wil,ghi,fri}
\begin{eqnarray}
K_{ff}&=&1+\left(13-2\pi\sqrt3\right)\hat\lambda
+\left[12-169\pi\sqrt3
+170\zeta(2)-252\zeta(3)+12\left(13\pi+19\sqrt3\right)\right.
\nonumber\\
&&{}\times\left.\cl\left(\frac{\pi}{3}\right)\right]\hat\lambda^2
\nonumber\\
&\approx&1+11.1\%\left(\frac{M_H}{\mbox{TeV}}\right)^2
-8.9\%\left(\frac{M_H}{\mbox{TeV}}\right)^4,
\nonumber\\
K_{VV}&=&1+\left[19-6\pi\sqrt3-10\zeta(2)\right]\hat\lambda
+62.0\,\hat\lambda^2
\nonumber\\
&\approx&1+14.6\%\left(\frac{M_H}{\mbox{TeV}}\right)^2
+16.9\%\left(\frac{M_H}{\mbox{TeV}}\right)^4,
\label{eq:gbet}
\end{eqnarray}
where $\hat\lambda=\lambda/(4\pi)^2$ and $\cl{}$ is Clausen's integral.
The $O(\lambda)$ terms in Eq.~(\ref{eq:gbet}) have been known for a long time
\cite{vel,wil}.
$K_{ff}$ and $K_{VV}$ are displayed as functions of $M_H$ in
Fig.~\ref{fig:frink}.
The $O(\lambda^2)$ terms of $K_{ff}$ and $K_{VV}$ start to exceed the
$O(\lambda)$ ones in magnitude at $M_H=1114$~GeV and 930~GeV, respectively.
These values mark a perturbative upper bound on $M_H$.
The nonperturbative value of $K_{VV}$ at $M_H=727$~GeV may be extracted from a
recent lattice simulation of elastic $\pi\pi$ scattering in the framework of
the four-dimensional O(4)-symmetric nonlinear $\sigma$ model in the broken
phase, where the $\sigma$ resonance was observed \cite{goe}.

\begin{figure}[ht]
\begin{center}
\begin{tabular}{cc}
\parbox{7.5cm}{\epsfig{file=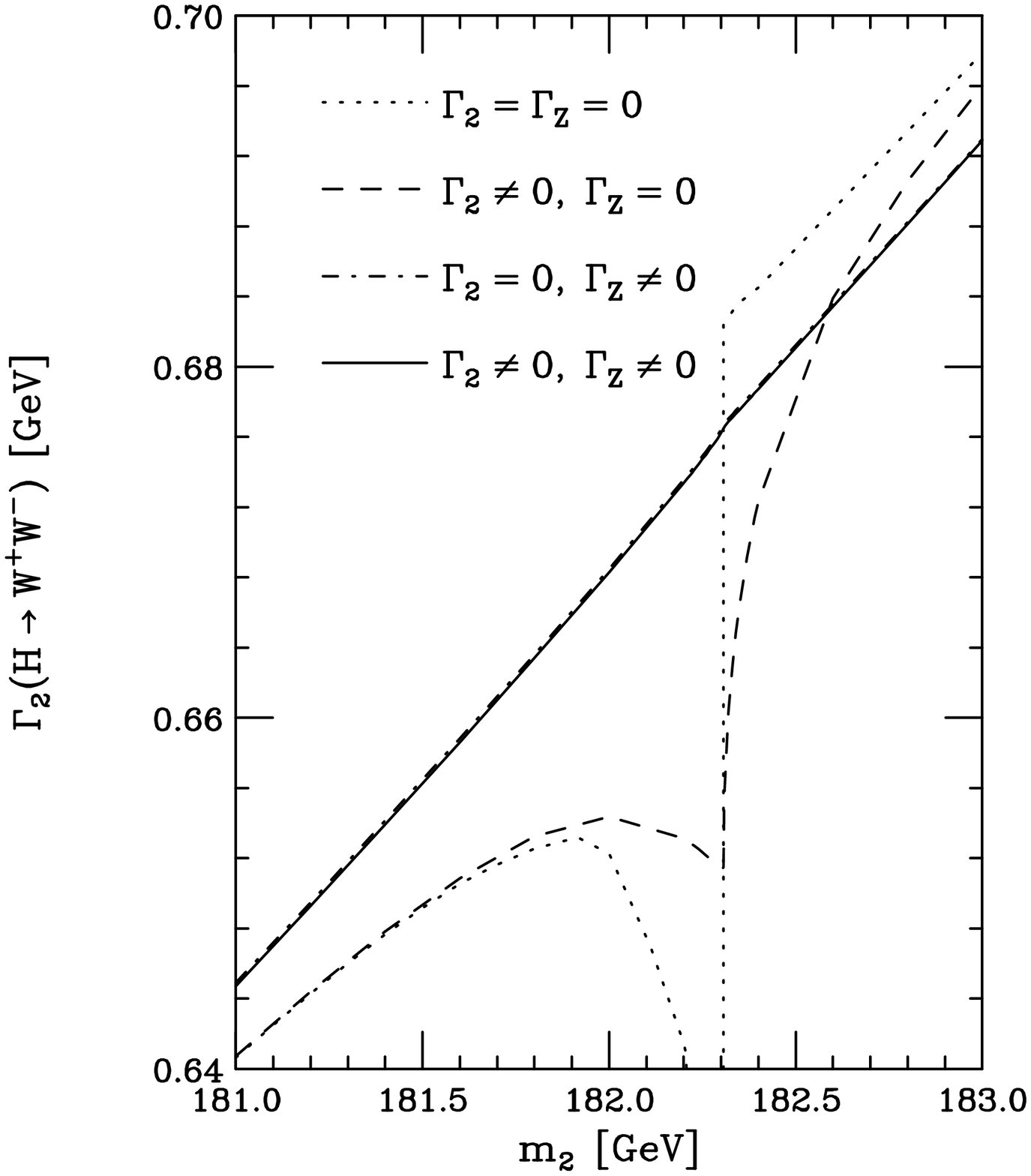,width=7.5cm}}
&
\parbox{7.5cm}{\epsfig{file=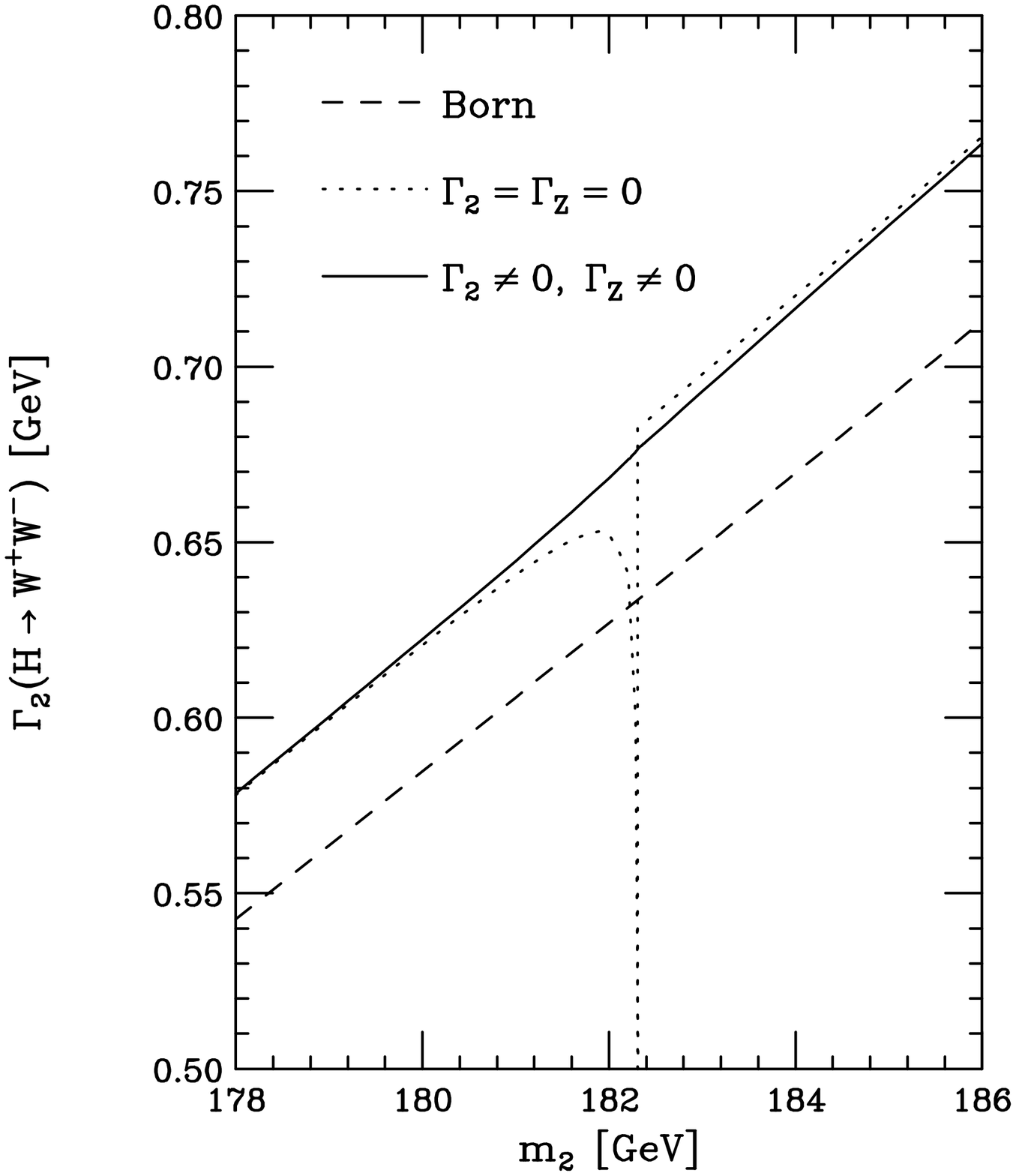,width=7.5cm}}\\
(a) & (b)
\end{tabular}
\caption{$\Gamma(H\to W^+W^-)$ as a function of $M_H$ \protect\cite{pal}.
(a) The threshold singularity at $M_H=2M_Z$ in the OS scheme (dotted line) is 
regularized by adopting the pole scheme (dashed line), allowing for the
$Z$-boson width to be finite (dot-dashed line), or both (solid line).
(b) The one-loop results in the OS scheme (dotted line) and in the pole
scheme with $\Gamma_Z\ne0$ (solid line) are compared with the tree-level 
result (dashed line).}
\label{fig:thr}
\end{center}
\end{figure}

At one loop in the conventional OS renormalization scheme, the production and
decay rates of the Higgs boson exhibit singularities proportional to
$(2M_V-M_H)^{-1/2}$ as $M_H$ approaches $2M_V$ from below \cite{hww}.
This problem is of phenomenological interest because the values $2M_W$ and
$2M_Z$, corresponding to the $W$- and $Z$-boson pair production thresholds,
lie within the $M_H$ range favoured by the arguments presented in
Sect.~\ref{sec:one}.
We recall that the OS mass $M$ and total decay width $\Gamma$ of an unstable
boson are defined as
\begin{eqnarray}
M^2&=&M_0^2+\re A(M^2),
\nonumber\\
M\Gamma&=&-Z\im A(M^2),
\nonumber\\
Z&=&\left[1-\re A^\prime(M^2)\right]^{-1},
\end{eqnarray}
where $M_0$ and $A(s)$ are the bare mass and unrenormalized self-energy, 
respectively, appearing in the propagator $i\left[s-M_0^2-A(s)\right]^{-1}$.
However, $A(s)$ possesses a branch point if $s$ is at a threshold.
If the threshold is due to a two-particle state with zero orbital angular
momentum, then $\re A^\prime(s)$ diverges as $1/\beta$, where $\beta$
is the relative velocity common to the two particles, as the threshold is
approached from below \cite{hww,bha}.
These threshold singularities are eliminated when the definitions of mass and
total decay width are based on the complex-valued position $\bar s$ of the
propagator's pole \cite{bha,pal}, as \cite{pal}
\begin{eqnarray}
\bar s&=&M_0^2+A(\bar s)=m_2^2-im_2\Gamma_2,
\nonumber\\
m_2\Gamma_2&=&-Z_2\im A\left(m_2^2\right),
\nonumber\\
Z_2&=&\left[1-\frac{\im A\left(m_2^2\right)-\im A(\bar s)}{m_2\Gamma_2}
\right]^{-1}.
\end{eqnarray}
This is illustrated in Figs.~\ref{fig:thr} (a) and (b) for
$\Gamma(H\to W^+W^-)$ in the vicinity of $M_H=2M_Z$.

\begin{figure}[ht]
\begin{center}
\centerline{\epsfig{file=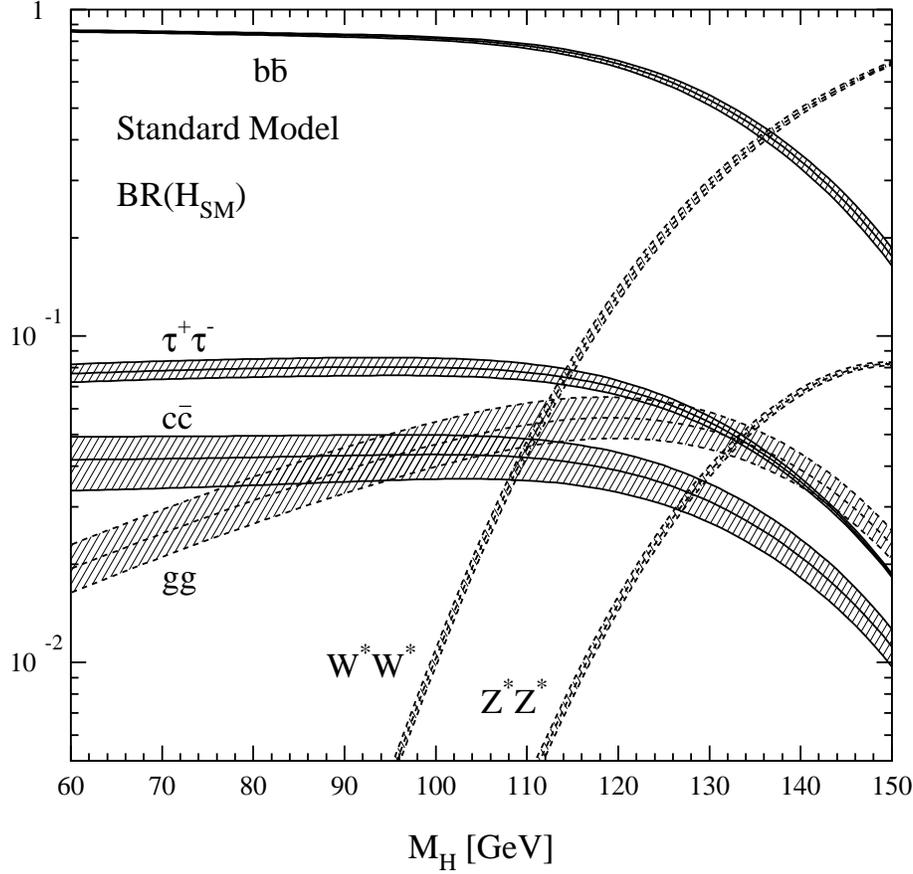,width=12cm,bbllx=8pt,bblly=264pt,bburx=573pt,%
bbury=812pt,clip=}}
\caption{Higgs-boson decay branching fractions, including theoretical
uncertainties, as functions of $M_H$ \protect\cite{dsz}.}
\label{fig:br}
\end{center}
\end{figure}

It is fair to say that radiative corrections for Higgs-boson decays have been
explored to a similar degree as those for $Z$-boson decays.
Unfortunately, this does not necessarily lead to similarly precise theoretical
predictions.
In fact, the errors on the latter are dominated by parametric uncertainties,
mainly by those in $\alpha_s^{(5)}(M_Z)$ and the quark masses (see
Fig.~\ref{fig:br}) \cite{gro,dsz}.

\boldmath
\section{\label{sec:three}Production in $e^+e^-$ collisions}
\unboldmath

\begin{figure}[ht]
\begin{center}
\centerline{
\epsfig{figure=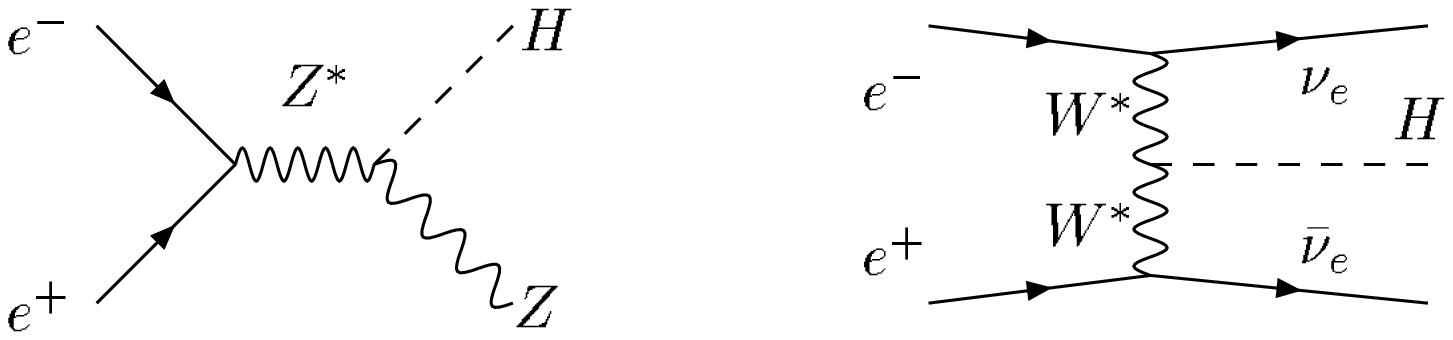,width=12cm,bbllx=84pt,bblly=620pt,bburx=503pt,%
bbury=718pt}}
\caption{Feynman diagrams for Higgs-strahlung and $W^+W^-$ fusion
\protect\cite{tdr}.}
\label{fig:fey}
\end{center}
\end{figure}

\begin{figure}[ht]
\begin{center}
\centerline{
\epsfig{figure=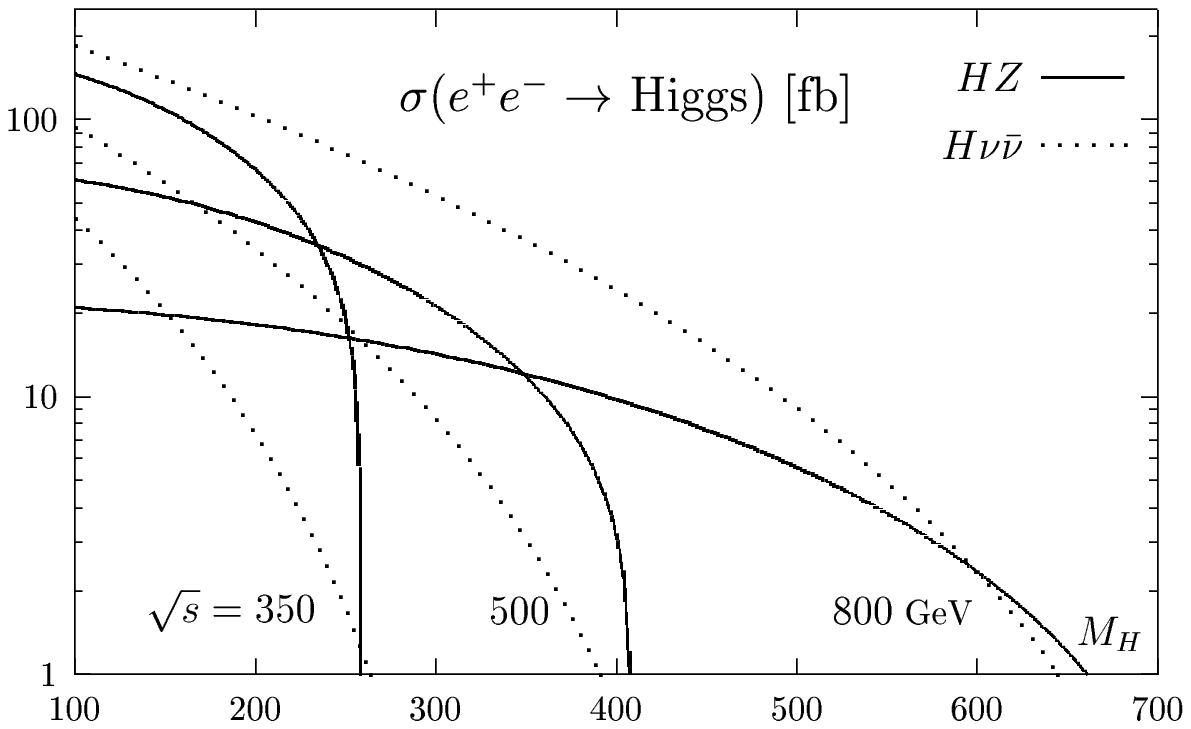,width=12cm,bbllx=137pt,bblly=485pt,bburx=475pt,%
bbury=702pt}}
\caption{Total cross sections of Higgs-strahlung and $W^+W^-$ fusion as 
functions of $M_H$ \protect\cite{tdr}.}
\label{fig:xs}
\end{center}
\end{figure}

The dominant mechanisms of Higgs-boson production in $e^+e^-$ collisions are
Higgs-strah\-lung and $W^+W^-$ fusion, which, at the tree level, proceed
through the Feynman diagrams depicted in Fig.~\ref{fig:fey}.
The cross section of $ZZ$ fusion, $e^+e^-\to e^+e^-H$, is approximately one
order of magnitude smaller than the one of $W^+W^-$ fusion, because of weaker
couplings.
The total cross section of Higgs-strahlung reads
\begin{equation}
\sigma(e^+e^-\to ZH)=\frac{g_{ZZH}^2}{4\pi}\,
\frac{G_F\left(v_e^2+a_e^2\right)}{96\sqrt2}\,
\frac{\sqrt\lambda\left(\lambda+12sM_Z^2\right)}{s^2D},
\end{equation}
where $v_f=2I_f-4s_w^2Q_f$ and $a_f=2I_f$ are the $Zf\bar f$ vector and
axial-vector couplings, respectively, $\sqrt s$ is the centre-of-mass energy,
$\lambda=\left[s-(M_Z+M_H)^2\right]\left[s-(M_Z-M_H)^2\right]$, and
$D=\left(s-M_Z^2\right)^2+M_Z^2\Gamma_Z^2$.
Here, $I_f$ is the third component of weak isospin of the left-handed
component of $f$, $Q_f$ is the electric charge of $f$, and
$c_w^2=1-s_w^2=M_W^2/M_Z^2$.
The one of $VV$ fusion may expressed as a one-dimensional integral \cite{ffzh}.
They are both shown in Fig.~\ref{fig:xs} as functions of $M_H$ for
$\sqrt s=350$, 500, and 800~GeV \cite{tdr}.

\begin{figure}[ht]
\begin{center}
\centerline{
\epsfig{figure=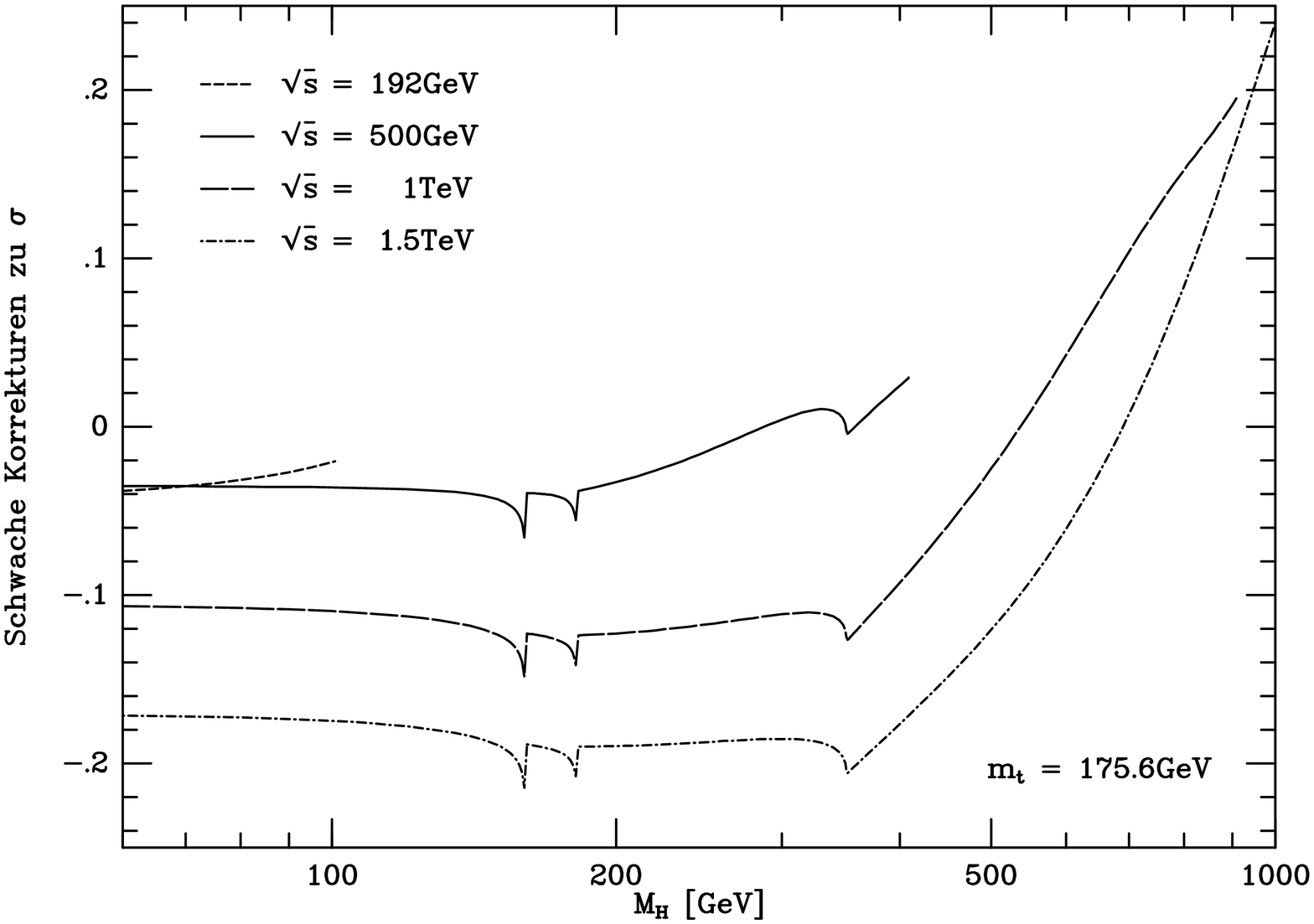,height=12cm,bbllx=27pt,bblly=54pt,bburx=532pt,%
bbury=765pt,angle=-90}}
\caption{Weak correction to $\sigma(e^+e^-\to ZH)$ as a function of $M_H$
\protect\cite{ffzh}.}
\label{fig:rc}
\end{center}
\end{figure}

As for the Higgs-strahlung process, the electromagnetic \cite{ffzh,ber} and
weak \cite{ffzh,den} corrections are fully known at one loop.
The latter is shown in Fig.~\ref{fig:rc} as a function of $M_H$ for
$\sqrt s=192$~GeV, 500~GeV, 1~TeV, and 2~TeV.

The electroweak corrections for $VV$ fusion, a $2\to3$ process, are not yet
available.
However, the leading effects can be conveniently included as follows.
The bulk of the initial-state bremsstrahlung can be taken into account in the
so-called leading logarithmic approximation provided by the structure-function
method, by convoluting the tree-level cross section with a radiator function,
which is known through $O(\alpha^2)$ and can be further improved by
soft-photon exponentiation \cite{bee}.
The residual dominant corrections of fermionic origin can be incorporated in a
systematic and convenient fashion by invoking the so-called improved Born
approximation (IBA) \cite{iba}.
These are contained in $\Delta\rho$ and
$\Delta\alpha=1-\alpha/\overline\alpha$, which parameterizes the running of
Sommerfeld's fine-structure constant from its value $\alpha$ defined in
Thomson scattering to its value $\overline\alpha$ measured at the $Z$-boson
scale.
The recipe is as follows.
Starting from the Born formula expressed in terms of $c_w$, $s_w$, and
$\alpha$, one substitutes
\begin{eqnarray}
\alpha&\to&\overline\alpha={\alpha\over1-\Delta\alpha},
\nonumber\\
c_w^2&\to&\overline c_w^2=1-\overline s_w^2=c_w^2(1-\Delta\rho).
\end{eqnarray}
One then eliminates $\overline\alpha$ in favour of $G_F$ by exploiting the
relation
\begin{equation}
{\sqrt2\over\pi}G_F={\overline\alpha\over\overline s_w^2M_W^2}
={\overline\alpha\over\overline c_w^2\overline s_w^2M_Z^2}(1-\Delta\rho),
\end{equation}
which correctly accounts for the leading fermionic corrections.
Finally, one includes the corrections enhanced by $X_t$ that are generated by
Eq.~(\ref{eq:eff}).
One thus obtains the correction factors
\begin{eqnarray}
K_{ffH}&=&1+2\delta_{ZZH}
+2\left[1-4c_w^2\left(\frac{Q_ev_e}{v_e^2+a_e^2}+\frac{Q_fv_f}{v_f^2+a_f^2}
\right)\right]\Delta\rho,
\nonumber\\
K_{\nu_e\nu_eH}&=&1+2\delta_{WWH},
\nonumber\\
K_{eeH}&\approx&K_{ffH},
\end{eqnarray}
for Higgs-strahlung, $W^+W^-$ fusion, and $ZZ$ fusion, respectively
\cite{fer}.
The interference of the scattering amplitudes for $\nu_e\bar\nu_eH$ and 
$e^+e^-H$ production by Higgs-strahlung with those for $W^+W^-$ and $ZZ$
fusion, respectively, is negligible for $\sqrt s>M_Z+M_H$ \cite{kil}.
It is important to keep in mind that the IBA is only reliable if
$\sqrt s,M_H\ll2M_t$.

It may be possible to operate a future $e^+e^-$ LC in $e^-e^-$, $e^\pm\gamma$,
or $\gamma\gamma$ modes.
In $e^-e^-$ collisions, Higgs bosons will be mainly produced via $ZZ$ fusion,
$e^-e^-\to e^-e^-H$ \cite{heu}.
Its cross section emerges from the one of $e^+e^-\to e^+e^-H$ by crossing
symmetry, as explained in Ref.~\cite{ffzh}, and it has a size very similar to
the latter.
The dominant Higgs-boson production mechanisms in $e^\pm\gamma$ collisions 
include the processes $e^\pm\gamma\to\nu_eW^\pm H$ \cite{boo,hwz},
$e^\pm\gamma\to e^\pm ZH$ \cite{boo}, and
$e^\pm\gamma\to e^\pm\gamma\gamma\to e^\pm H$ \cite{ebo}, which proceeds via
charged-fermion and $W$-boson loops. 
In $\gamma\gamma$ collisions, Higgs bosons will be chiefly created through
$\gamma\gamma$ fusion, $\gamma\gamma\to H$ \cite{tel}, which is mediated by
the same types of loops.
Cutting open the $W$-boson loops leads to the process
$\gamma\gamma\to W^+W^-H$ \cite{bai}, which benefits from the huge cross
section of the parent process $\gamma\gamma\to W^+W^-$.
The process $\gamma\gamma\to t\bar tH$ \cite{gin} is sensitive to the top 
Yukawa coupling $g_{ttH}$, but it suffers from phase-space suppression.

\section{\label{sec:four}Quantum numbers and couplings from final states}

The spin, parity, and charge-conjugation quantum numbers $J^{PC}$ of Higgs
bosons can be determined at a future $e^+e^-$ LC in a model-independent way.
The observation of the decay or fusion processes
$H\rightleftharpoons\gamma\gamma$ would rule out $J=1$ by the Landau-Yang
theorem and, at the same time, fix $C$ to be positive \cite{ver}.

The angular distribution of $e^+e^-\to ZH$ depends on $J$ and $P$.
The SM Higgs boson is a $0^{++}$ state, and its couplings to two $Z$ bosons is
proportional to $\vec\epsilon\cdot\vec{\epsilon^\prime}$ in the laboratory
frame, where $\vec\epsilon$ and $\vec{\epsilon^\prime}$ are the polarization
three-vectors of the $Z$ bosons.
In order to distinguish the SM Higgs boson from a $CP$-odd $0^{-+}$ state $A$,
or a $CP$-violating mixture of the two, which will be generically denoted by
$\Phi$, one may consider a $ZZ\Phi$ coupling of the form \cite{DK}
\begin{equation}
\Gamma_{ZZ\Phi}=ig_{ZZH}\left(g^{\mu\mu^\prime}+i\frac{\eta}{M_Z^2}
\epsilon^{\mu\mu^\prime\nu\nu^\prime}p_\nu p_{\nu^\prime}^\prime\right),
\label{eq:zzp}
\end{equation}
where $p,p^\prime$ and $\mu,\mu^\prime$ are the incoming four-momenta and
Lorentz indices of the two $Z$ bosons, respectively, and $\eta$ is a
dimensionless factor.
In the case $\eta=0$, we recover the SM Higgs boson, while the absence of the
first term in Eq.~(\ref{eq:zzp}) corresponds to an $A$ boson.
In the laboratory frame, the $ZZA$ coupling is proportional to
$(\vec\epsilon\times\vec{\epsilon^\prime})\cdot(\vec p-\vec{p^\prime})$.
The $CP$-odd case is realized in the minimal supersymmetric extension of the
SM and in two-Higgs-doublet models (2HDM) without $CP$ violation, in which the
$ZZA$ couplings are induced at the level of quantum loops.
However, in a more general scenario, $\eta$ need not be loop suppressed, and
it is useful to allow for $\eta$ to be arbitrary in the experimental data 
analysis.
In a general 2HDM, the three neutral Higgs bosons correspond to arbitrary
mixtures of $CP$ eigenstates, and their production and decay processes exhibit
$CP$ violation.
The differential cross section of $e^+e^-\to Z\Phi$ that results from the
coupling of Eq.~(\ref{eq:zzp}) reads
\begin{eqnarray}
\frac{d\sigma(e^+e^-\to Z\Phi)}{d\cos\theta}&=&\frac{g_{ZZH}^2}{4\pi}\,
\frac{G_F\left(v_e^2+a_e^2\right)}{16\sqrt2}\,\frac{M_Z^2\sqrt\lambda}{sD}
\left[1+\frac{\lambda}{8sM_Z^2}\sin^2\theta
+\eta\frac{2v_ea_e}{v_e^2+a_e^2}\,\frac{\sqrt\lambda}{M_Z^2}\cos\theta\right.
\nonumber\\
&&+\left.\eta^2\frac{\lambda}{8M_Z^4}(1+\cos^2\theta)\right],
\label{eq:cos}
\end{eqnarray}
where $\theta$ is the polar angle of the $Z$ boson w.r.t.\ to the beam axis
in the laboratory frame.
Thus, the angular distribution of $e^+e^-\to ZA$, namely
$(1/\sigma)d\sigma(e^+e^-\to ZA)/d\cos\theta=(3/8)(1+\cos^2\theta)$, is very
distinct from the SM one, which is
$(1/\sigma)d\sigma(e^+e^-\to ZH)/d\cos\theta\approx(3/4)\sin^2\theta$ for
$\sqrt s\gg M_Z$ \cite{ver}.
The presence of the interference term (linear in $\eta$) in
Eq.~(\ref{eq:cos}), would generate a forward-backward asymmetry, which would
be a clear signal for $CP$ violation.
Another discriminator between the $CP$-even and $CP$-odd cases is provided by
the threshold behaviour of the cross section, which is proportional to
$\beta=\sqrt\lambda/s$ and $\beta^3$, respectively \cite{ver}.
In the most general situation, where the particle produced in association with
the $Z$ boson corresponds to a $J^P$ state, the threshold behaviour is 
$\sigma(e^+e^-\to Z\Phi)\propto\beta^n$, where $n$ is listed in
Table~\ref{tab:jp} \cite{mil}.
We conclude that the observation of a threshold behaviour linear in $\beta$
would rule out the assignments $J^P=0^-,1^-,2^-,3^\pm,4^\pm,\ldots$.
\begin{table}
\begin{center}
\caption{Values of $n$ characterizing the threshold behaviour of
$\sigma(e^+e^-\to Z\Phi)$, where $\Phi$ is a $J^P$ state.}
\label{tab:jp}
\smallskip
\begin{tabular}{|c|c|}
\hline\hline
$J^P$ & $n$ \\
\hline
$0^+$, $1^+$, $2^+$ & 1 \\
$0^-$, $1^-$, $2^-$ & 3 \\
$3^-$, $4^+$, $5^-$, \dots & $2J-3$ \\
$3^+$, $4^-$, $5^+$, \dots & $2J-1$ \\
\hline\hline
\end{tabular}
\end{center}
\end{table}

\begin{figure}[ht]
\begin{center}
\centerline{\epsfig{figure=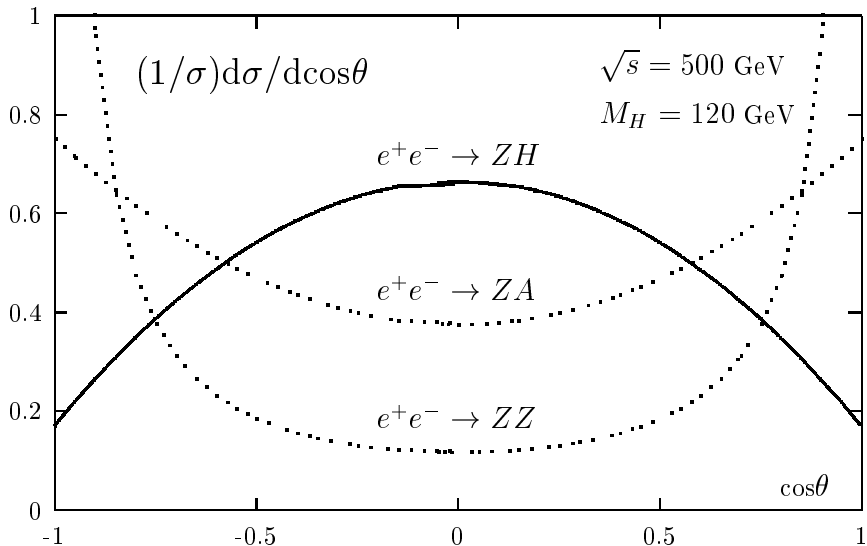,width=12cm}}
\caption{$\sigma(e^+e^-\to ZH)$, $\sigma(e^+e^-\to ZA)$, and
$\sigma(e^+e^-\to ZZ)$ as functions of $\cos\theta$ \protect\cite{ver}.}
\label{fig:za}
\end{center}
\end{figure}

The angular distribution of $e^+e^-\to Z\Phi$ can also be exploited to
establish the $J=0$ nature of the Higgs bosons.
To this end, it should be compared with the one of $e^+e^-\to ZZ$, which 
exhibits a distinctly different angular momentum structure.
Owing to the electron exchange in the $t$-channel, the $e^+e^-\to ZZ$
amplitude is built up by many partial waves, which peak in the forward and
backward directions.
In Fig.~\ref{fig:za}, the angular distributions of $ZH$, $ZA$, and $ZZ$
production are shown for $\sqrt s=500$~GeV, assuming a Higgs-boson mass of
120~GeV.
 
The angular distribution of the $Z\to f\bar f$ decay products of the secondary
$Z$ boson in the Higgs-strahlung process will also help us to distinguish a
$CP$-even Higgs boson from a $CP$-odd one or a spin-one boson.
In fact, at high energies, the $Z$ bosons from $e^+e^-\to ZH$ are dominantly
longitudinally polarized, while the ones from $e^+e^-\to ZA$ and
$e^+e^-\to ZZ$ are fully and dominantly transversely polarized, respectively
\cite{ver}.
Calling the polar angle enclosed between the flight direction of the decay 
fermion $f$ in the $Z$-boson rest frame and the $Z$-boson flight direction in
the laboratory frame $\theta^*$ and the corresponding azimuthal angle w.r.t.\
the plane spanned by the beam axis and the $Z$-boson flight direction 
$\phi^*$, longitudinal [transverse] $Z$ bosons lead to a distribution
proportional to $\sin^2\theta^*$ [$(1\pm\cos\theta^*)^2$] after integrating
over $\phi^*$ \cite{ver}.
On the other hand, after integrating over $\theta$ and $\theta^*$, we have
\begin{eqnarray}
\frac{d\sigma(e^+e^-\to ZH)}{d\phi^*}
&\propto&1+a_1\cos\phi^*+a_2\cos(2\phi^*),
\nonumber\\
\frac{d\sigma(e^+e^-\to ZA)}{d\phi^*}&\propto&1-\frac{1}{4}\cos(2\phi^*),
\end{eqnarray}
where the coefficients $a_i$ depend on $\sqrt s$ and $f$ \cite{ver}.
The distribution of
$e^+e^-\to Z\Phi\to\left(f\bar f\right)\left(f^\prime\bar f^\prime\right)$ in
$\theta$, $\theta^*$, and $\phi^*$ for the coupling of Eq.~(\ref{eq:zzp}) may
be found in Ref.~\cite{DK}.

The determination of the $J^{PC}$ quantum numbers of the Higgs bosons can be 
refined by taking the angular distributions of their decay products into
account.
The $J=0$ property manifests itself in the complete absence of angular 
correlations between the initial- and final-state particles.
The criteria to distinguish between $CP$-even and $CP$-odd Higgs bosons or 
mixtures thereof include the polarization of the vector bosons $V=W,Z$ in the
decay $\Phi\to VV$, the distribution in the mass $M_*$ of the virtual boson
$V^*$ in the decay $\Phi\to VV^*$, and characteristic features of the
angular distribution of the decay
$\Phi\to V^*V^*\to\left(f\bar f\right)\left(f^\prime\bar f^\prime\right)$
\cite{ver,DK}.

In the effective-Lagrangian approach, the $ZZ\Phi$ coupling in
Eq.~(\ref{eq:zzp}) is not the most general one \cite{sto,hag,grz}.
In fact, the first term may come with a fudge factor different from unity, and
there may be two more independent $CP$-even terms.
Similarly, there may be an effective $Z\gamma\Phi$ coupling, involving two
$CP$-even and one $CP$-odd terms.
The most general effective $ZV\Phi$ interaction Lagrangian reads
\cite{sto,hag}
\begin{eqnarray}
{\cal L}_{\rm eff}&=&\frac{g_{ZZH}}{2}(1+a_Z)HZ_\mu Z^\mu
+\frac{g_{ZZH}}{M_Z^2}\sum_{V=Z,\gamma}\left[b_VHZ_{\mu\nu}V^{\mu\nu}+
c_V\left(\partial_\mu HZ_\nu-\partial_\nu HZ_\mu\right)V^{\mu\nu}\right.
\nonumber\\
&&{}+\left.\tilde{b}_VHZ_{\mu\nu}\tilde{V}^{\mu\nu}\right],
\end{eqnarray}
where $V_{\mu\nu}=\partial_\mu V_\nu-\partial_\nu V_\mu$ and
$\tilde{V}_{\mu\nu}=\epsilon_{\mu\nu\alpha\beta}V^{\alpha\beta}$.
Here, we have neglected the scalar components of the vector bosons, by putting
$\partial_\mu Z^\mu=\partial_\mu V^\mu=0$.
The couplings $a_Z$, $b_Z$, $c_Z$, $b_\gamma$, and $c_\gamma$ are $CP$-even,
while $\tilde b_Z$ and $\tilde b_\gamma$ are $CP$-odd.

\begin{figure}[ht]
\begin{center}
\begin{tabular}{cc}
\parbox{7.5cm}{\epsfig{file=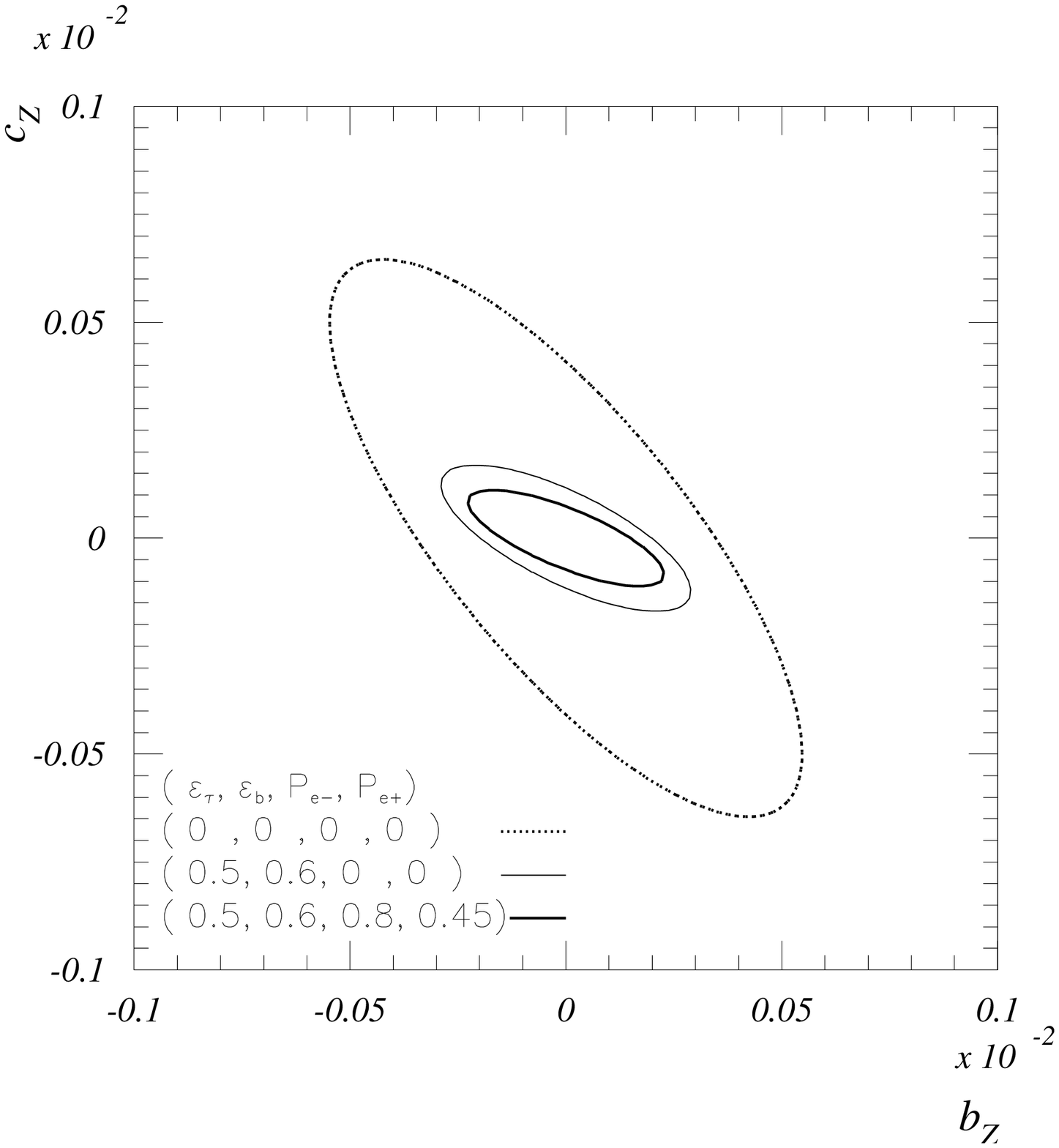,width=7.5cm}}
&
\parbox{7.5cm}{\epsfig{file=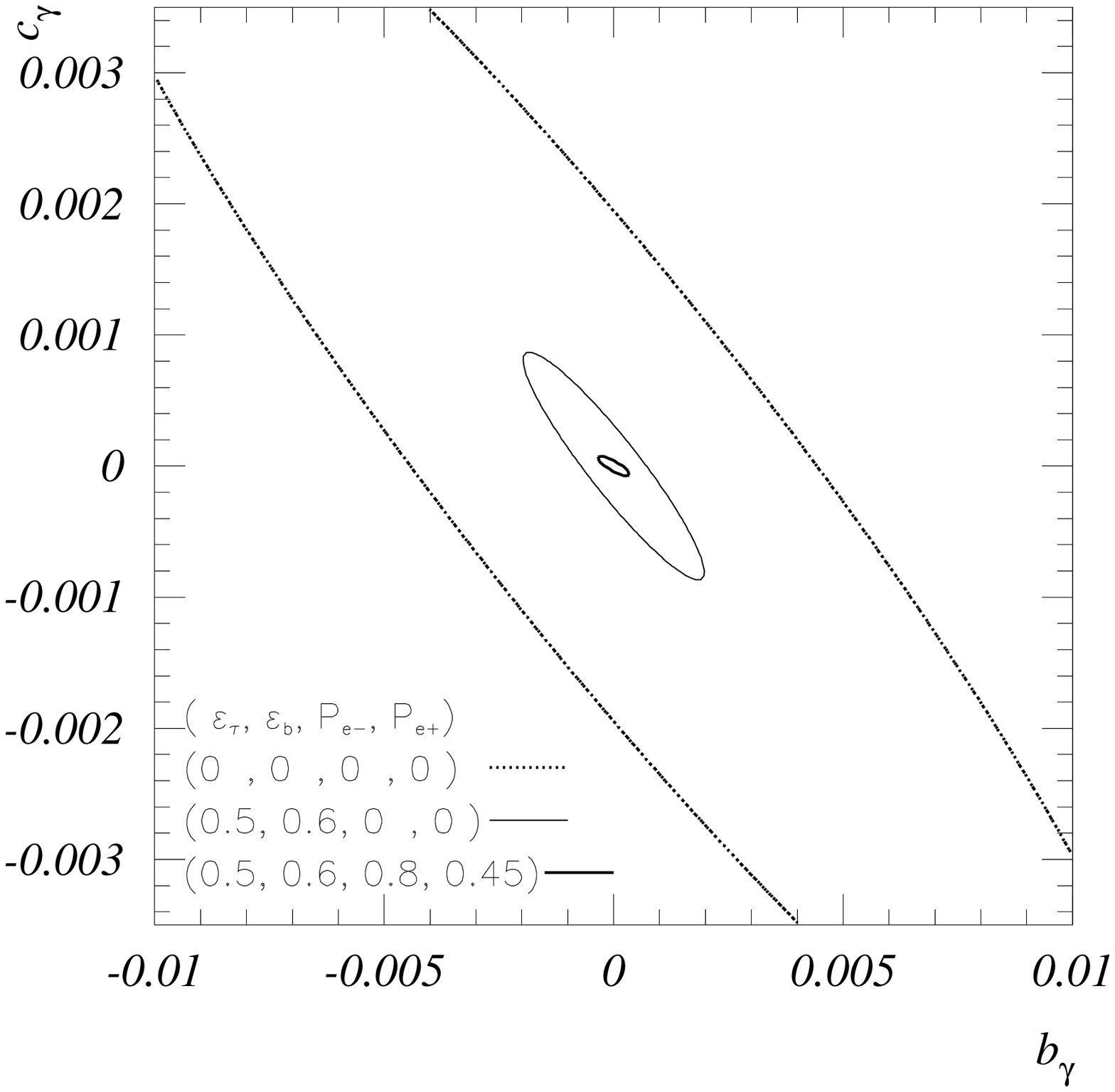,width=7.5cm}}\\
(a) & (b)
\end{tabular}
\caption{Contours of $\chi^2=1$ in the (a) $(b_Z,c_Z)$ and (b)
$(b_\gamma,c_\gamma)$ planes for 300~fb$^{-1}$ of data at
$\protect\sqrt s=500$~GeV \protect\cite{hag}.
In each case, the other degrees of freedom are integrated out.}
\label{fig:hag}
\end{center}
\end{figure}

With sufficiently high luminosity, it should be possible to determine, by 
means of the optimal-observable method \cite{oom,gun}, most of these couplings
from the angular distribution of $e^+e^-\to Z\Phi\to\left(f\bar f\right)\Phi$.
The achievable bounds can be significantly improved by measuring the 
tau-lepton helicities, identifying the bottom-hadron charges, polarizing the
electron and positron beams, and running at two different values of $\sqrt s$
\cite{hag}.
The results for energy $\sqrt s=500$~GeV, luminosity $L=300$~fb$^{-1}$,
efficiencies $\epsilon_\tau=50\%$ and $\epsilon_b=60\%$, and polarizations
$P_{e^-}=\pm80\%$ and $P_{e^+}=\mp45\%$ are summarized in 
Table~\ref{tab:hag} and visualized in Figs.~\ref{fig:hag}(a) and (b).
Here, the couplings are assumed to be real, and $a_Z$ is fixed.
In order to also determine $a_Z$, one needs to perform the experiment at two
different values of $\sqrt s$.
We observe that the $ZZ\Phi$ couplings are generally well constrained, even
for $\epsilon_\tau=\epsilon_b=P_{e^-}=P_{e^+}=0$, while the $Z\gamma\Phi$
couplings are not.
The constraints on the latter may be significantly improved by the above-named
options, especially by beam polarization.
\begin{table}[ht]
\begin{center}
\caption{Optimal errors on general $ZZ\Phi$ and $Z\gamma\Phi$ couplings for
300~fb$^{-1}$ of data at $\protect\sqrt s=500$~GeV \protect\cite{hag}.}
\label{tab:hag}
\smallskip
\begin{tabular}{|c|ccc|}
\hline\hline
$\epsilon_\tau$ & --- & 0.5 & 0.5 \\
$\epsilon_b$    & --- & 0.6 & 0.6 \\
$|P_{e^-}|$     & --- & --- & 0.8 \\
$|P_{e^+}|$     & --- & --- & 0.6 \\
\hline
$b_Z$ & 0.00055 & 0.00029 & 0.00023 \\
$c_Z$ & 0.00065 & 0.00017 & 0.00011 \\
$b_\gamma$ & 0.01232 & 0.00199 & 0.00036 \\
$c_\gamma$ & 0.00542 & 0.00087 & 0.00008 \\
$\tilde b_Z$ & 0.00104 & 0.00097 & 0.00055 \\
$\tilde b_\gamma$ & 0.00618 & 0.00101 & 0.00067 \\
\hline\hline
\end{tabular}
\end{center}
\end{table}

\begin{figure}[ht]
\begin{center}
\centerline{
\epsfig{figure=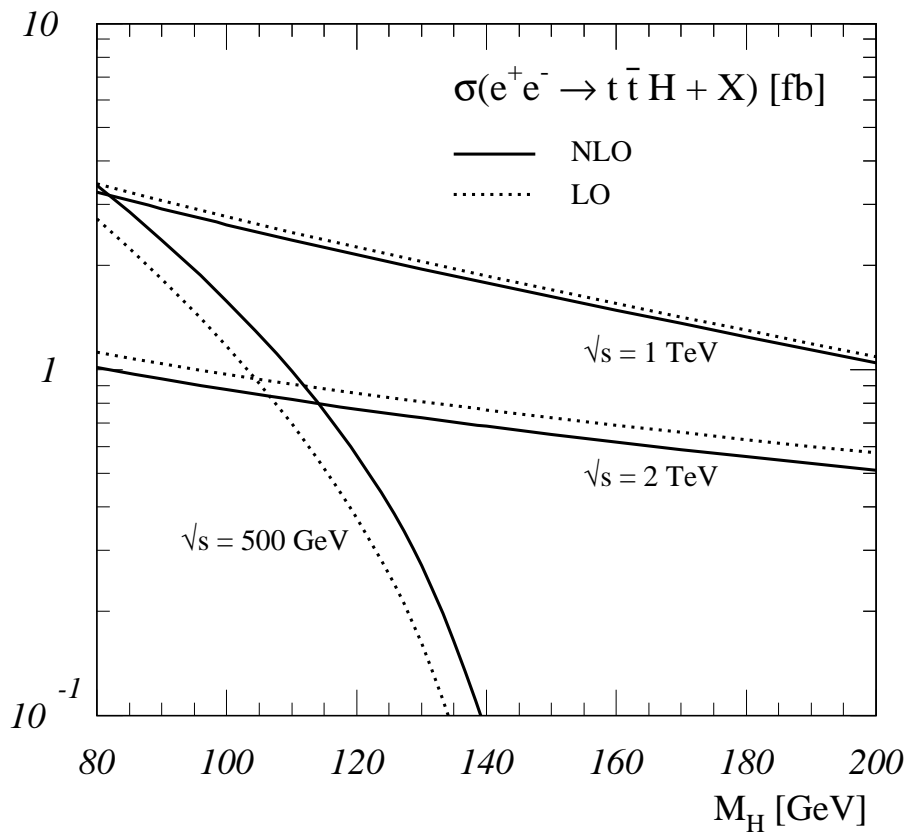,width=12cm,bbllx=26pt,bblly=16pt,bburx=323pt,%
bbury=290pt}}
\caption{$\sigma(e^+e^-\to t\bar tH)$ with and without QCD corrections as a
function of $M_H$ \protect\cite{dit}.}
\label{fig:tth}
\end{center}
\end{figure}

Once $g_{ZZH}$ has been pinned down, the top Yukawa coupling $g_{ttH}$ can be
extracted by studying the process $e^+e^-\to t\bar tH$ \cite{kal}.
The QCD correction to its cross section can be of either sign, depending on
$\sqrt s$, and reach a magnitude of several ten percent \cite{daw,dit}.
This may be seen from Fig.~\ref{fig:tth}, where the Born and QCD-corrected
cross sections are shown as functions of $M_H$ for $\sqrt s=500$~GeV, 1~TeV,
and 2~TeV.
Anomalous top Yukawa couplings may be extracted from the angular distribution
of $e^+e^-\to t\bar t\Phi$ with the help of the optimal-observable method
\cite{gun}.

\begin{figure}[ht]
\begin{center}
\begin{tabular}{cc}
\parbox{7.5cm}{
\epsfig{file=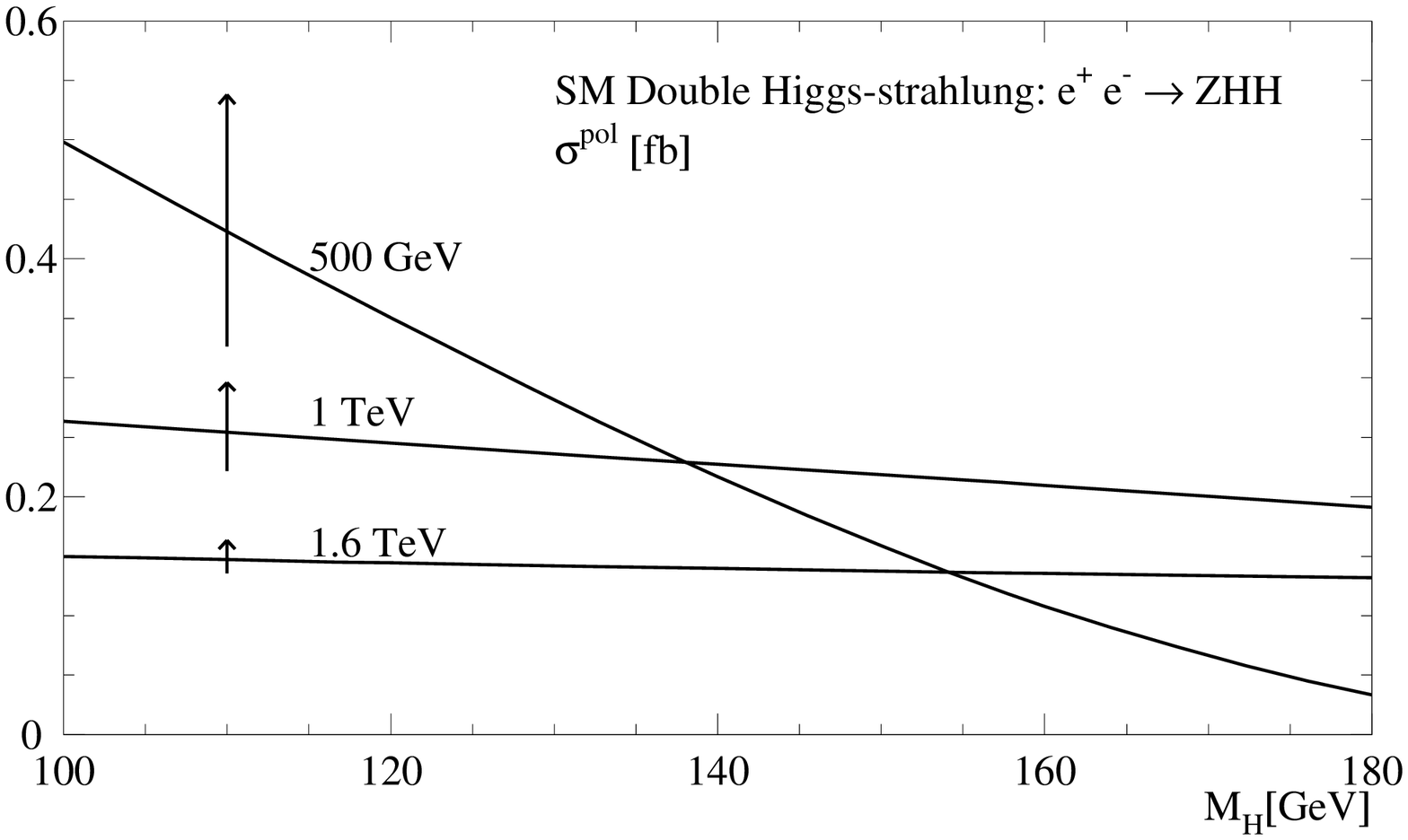,width=7.5cm,bbllx=0pt,bblly=0pt,bburx=567pt,%
bbury=340pt}}
&
\parbox{7.5cm}{
\epsfig{file=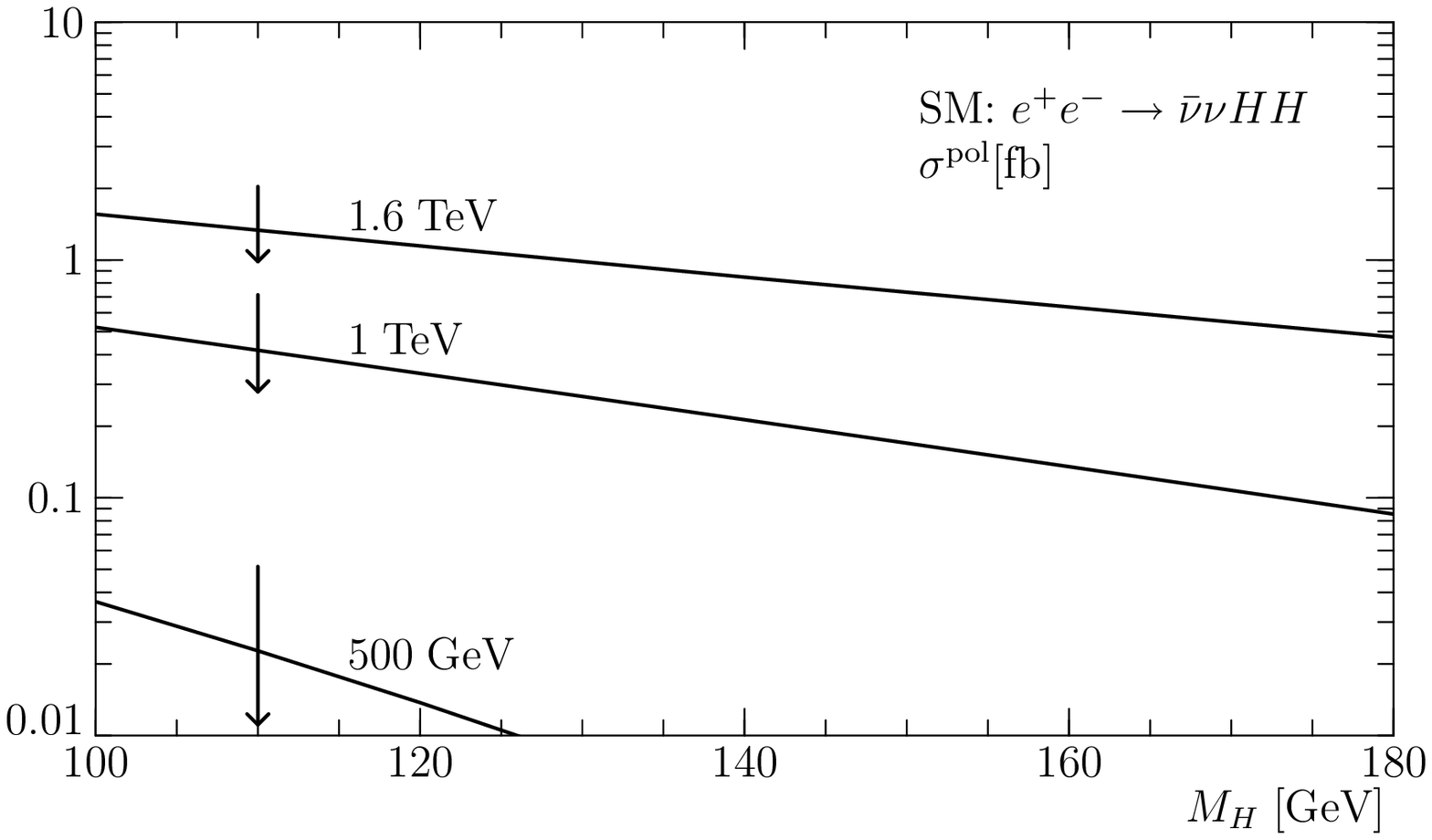,width=7.5cm,bbllx=52pt,bblly=455pt,bburx=538pt,%
bbury=741pt}}\\
(a) & (b)
\end{tabular}
\caption{(a) $\sigma(e^+e^-\to ZHH)$ and (b)
$\sigma(e^+e^-\to\nu_e\bar\nu_eHH)$ as functions of $M_H$ \protect\cite{mmm}.
The vertical arrows indicate the shifts in cross section induced by the
variation of $g_{HHH}$ by $\pm50\%$.}
\label{fig:mmm}
\end{center}
\end{figure}

The analysis of double Higgs-strahlung, $e^+e^-\to ZHH$, and $W^+W^-$ 
double-Higgs fusion, $e^+e^-\to\bar\nu_e\nu_eHH$, offers the possibility to
extract the trilinear Higgs self-coupling $g_{HHH}$ \cite{mmm}.
The cross sections of these two processes are relatively modest, but they can 
be enhanced by factors 2 and 4, respectively, by using beam polarization.
They are shown as functions of $M_H$ for $\sqrt s=500$~GeV, 1~TeV,
and 1.6~TeV in Figs.~\ref{fig:mmm}(a) and (b), respectively.
The sensitivity to $g_{HHH}$ is strongest close to the production thresholds.

\section{\label{sec:five}Conclusions and outlook}

We reviewed theoretical results that are relevant for the phenomenology of the
SM Higgs boson at a future $e^+e^-$ LC, putting special emphasis on radiative
corrections to its partial decay widths and production cross sections, and on
the logistics of extracting its quantum numbers and couplings from the
analysis of appropriate final states.
It is fair to say that theoretical predictions for partial decay widths and
production cross sections are generally in good shape.
However, the precision on the partial decay widths is limited by parametric
uncertainties, mainly by those in $\alpha_s^{(5)}(M_Z)$ and the quark masses.
The strategies for the determination of the Higgs profile are also well
elaborated.

The list of urgent tasks left to be done includes the calculation of the full
$O(\alpha)$ corrections for important $2\to3$ processes, such as $W^+W^-$ 
fusion, $ZZ$ fusion, and $t\bar tH$ associated production, and the inclusion
of background processes and detector simulation.

\vspace{1cm}
\begin{center}
{\bf Acknowledgements}
\end{center}
\smallskip

We thank Matthias Steinhauser for carefully reading this manuscript.
This work was supported in part by the Deutsche Forschungsgemeinschaft through
Grant No.\ KN~365/1-1, by the Bundesministerium f\"ur Bildung und Forschung
through Grant No.\ 05~HT1GUA/4, and by the European Commission through the
Research Training Network {\it Quantum Chromodynamics and the Deep Structure
of Elementary Particles} under Contract No.\ ERBFMRX-CT98-0194.

\end{document}